\documentclass[reprint,
 amsmath,amssymb,
 aps, 
 floatfix
]{revtex4-2}
\usepackage{caption}
\usepackage{nicefrac}
\usepackage[colorlinks=true,
linkcolor=black,                
citecolor=blue,                
filecolor=black,               
urlcolor=black,                
]{hyperref}
\usepackage{bbold}
\usepackage{placeins}
\usepackage{subcaption}
\usepackage{multirow}
\usepackage{graphicx}
\usepackage{dcolumn}
\usepackage{bm}
\usepackage{blkarray}
\usepackage{booktabs}
\usepackage{calc}
\newcolumntype{C}[1]{>{\centering\arraybackslash$}m{#1}<{$}}
\newlength{\mycolwd}                                         
\begin{document}

\preprint{APS/123-QED}

\title{A multi-type branching process method for modelling complex contagion on clustered networks}

\author{Leah A. Keating}
 \email{leah.keating@ul.ie}
\author{James P. Gleeson}%
\author{David J.P. O'Sullivan}
\affiliation{%
 MACSI, Department of Mathematics and Statistics, University of Limerick, Limerick V94 T9PX, Ireland
}%

\date{\today}

\begin{abstract}
Complex contagion adoption dynamics are characterised by a node being more likely to adopt after multiple network neighbours have adopted. We show how to construct multi-type branching processes to approximate complex contagion adoption dynamics on networks with clique-based clustering.  This involves tracking the evolution of a cascade via different classes of clique motifs that account for the different numbers of active, inactive and removed nodes. This discrete-time model assumes that active nodes become immediately and certainly removed in the next time step. This description allows for extensive Monte Carlo simulations (which are faster than network-based simulations), accurate analytical calculation of cascade sizes, determination of critical behaviour and other quantities of interest.
\end{abstract}

\maketitle


\section{Introduction}

Intuitively, we understand that the more social influence a person receives, the more likely they are to adopt a new behaviour. Examples of behaviours that have been shown to spread in this way are the use of Skype add-ons \cite{karsai2014}, the use of politically controversial hashtags on Twitter \cite{romero2011differences} and online fads \cite{Sprague2017}. Centola \cite{Centola2010} found that the adoption of health behaviour online was more likely after multiple friends of a user had already adopted the behaviour. This type of spread is known as a \textit{complex contagion}. In a complex contagion, given that an individual has not already adopted, the more network neighbours that have previously adopted, the more likely it is that an individual will adopt the next time they are exposed to the behaviour. Simple contagion dynamics, where the effects of multiple exposures are independent of each other \cite{borge2013cascading, Porter2016DynamicalNetworks}, are a natural model for disease spread where a virus is transmitted from infected to susceptible organisms because successful transmission is independent of the number of previous contacts with infected organisms. Simple contagion dynamics are well understood and can provide accurate models of some social spreading processes; however, this is not true in all cases \cite{Weng2013ViralityNetworks}. For example, Romero et al.~\cite{romero2011differences} tracked the spread of hashtags on Twitter and found that while some topics spread like a simple contagion, others exhibit complex contagion dynamics.\par
Social networks are known to have high levels of clustering or transitivity \cite{newman_social_2003, watts1998collective}, this is the extent to which ``a friend of my friend is a friend of mine." Clustering has been shown to inhibit diffusion in a network for simple contagions \cite{shirley2005}, it is understood that this is caused by having more edges shared by infected nodes in clustered networks, these edges are said to be redundant because they do not assist the diffusion. Centola's findings showed empirically that the spread of behaviour does not always follow this rule. \par
Branching processes are discrete-time stochastic processes which have been used to model infectious diseases in heterogeneous populations \cite{vazquez06, vazquez21} and diffusion cascades of information in networks \cite{gleeson2014competition, gleeson2016effects, Gleeson2014, gleeson2017avalanches, Aragon2017GenerativeChallenges, McSweeney2020}. Gleeson et al.~\cite{Gleeson2020a} derive analytical results for branching processes including cascade size, expected tree depth and structural virality \cite{goel2016structural}; the branching process theory allows not only the expectation but also the distribution of these metrics to be calculated. Simple branching processes cannot account for clustering. McSweeney \cite{McSweeney2020} extended the work of Hackett et al.~\cite{Hackett2011} to model complex contagion on a class of clustered networks, composed of triangles and single links, developed separately by Newman \cite{Newman2009} and Miller \cite{Miller2009} using two-type branching processes. In this paper, we will refer to this class of networks as Newman-Miller networks. The complex contagion dynamics in McSweeney's model are governed by a threshold model \cite{watts2002}. In this paper we extend McSweeney's work \cite{McSweeney2020} beyond two-type branching processes to account for the presence of higher-order cliques. Such higher-order structures have been accounted for under continuous-time dynamics, O'Sullivan et al.~\cite{OSullivan2015} adapted the method of Hébert-Dufrense et al.~\cite{Hebert-Dufresne2010} to model complex contagion on clustered networks using a SI compartmental model, which yielded results that provided analytical backing to Centola's experimental results \cite{Centola2010}. They studied the early time spreading behaviour for the contagion on networks with varying levels of clustering and under different amounts of social reinforcement. In this paper, we study complex contagion on clustered networks in discrete time using branching processes. This allows us to not only study the early-time behaviour as in \cite{OSullivan2015} but also to study the full cascades from birth to possible extinction.\par
Simple contagions on locally tree-like networks can be accurately modelled using simple branching processes \cite{Gleeson2020a}; however, these models do not allow for clustering which is an important driver of complex contagion \cite{Centola2010, OSullivan2015}. If we extend these models to a multi-type branching process (MTBP), we can capture the effects of clustering in a network. MTBPs have been used to model population dynamics in ecology since the mid 1940s; for a very accessible introduction see \cite{Caswell2018MatrixModels} and references therein. These models can track a population over time by dividing it into different life-stage compartments, allowing for properties such as varying fertility across ages to be captured. We borrow from the MTBP literature in ecology to develop a novel method of finding analytic results for complex contagion in networks with clique structures. We track clique motifs in the network where each motif type has a specific number of active, inactive and removed nodes. In a clique each node is connected to every other node; see Figs.~\ref{fig:schematic} and \ref{fig:4cl_motifs} for examples of clique motifs. The clique motifs in the cascade MTBP model are analogous to the life stages used to model populations in ecology. The MTBP begins with a seed node (or nodes) and a predefined number of motifs of different types (see Fig.~\ref{fig:ICs}), the process evolves by these motifs producing other motif types as their offspring according to a probability distribution.\par
There are two main choices made in defining our model, the first is the choice of complex contagion probability function and the other is the decision to use a MTBP. Complex contagion is commonly modelled using threshold models \cite{OSullivan2015, monsted2017, Unicomb2018ThresholdNetworks}, the Watts threshold model \cite{watts2002} was a seminal model of this class. In Watts' model a node becomes active if more than a specific fraction of its neighbours are active; however, Centola \cite{Centola2010} found that instead of exhibiting the binary behaviour assumed in threshold models, users' probability of adopting the health behaviour increased with each additional adopting neighbour. Here, we introduce a model for complex contagion which is an extension of the independent cascade model (ICM) \cite{kempe2003} and aims to capture the effect of additional exposures on the adoption probability as observed by Centola. Under these dynamics, in the social network context, given that an individual has not yet adopted, the more friends who have previously adopted the behaviour the greater the chance the individual will adopt the behaviour on the next exposure.\par
The rest of this paper is structured as follows. In Sec.~\ref{sec:cc_model} we describe the adoption probabilities associated with the complex contagion model, where given that adoption has not occurred, the more previous exposures received the greater the probability of adoption after the next exposure. In Sec.~\ref{sec:mtbp} we define the framework for the MTBP on 6-regular networks, we show how the transition probabilities are found and present the mean matrix, $\textbf{M}$, which we use to derive average properties of the cascades. The MTBP methodology can be extended to network distributions beyond the simple examples given in this section. In Sec.~\ref{sec:results} we give results from the MTBP, in particular, we identify the regions of parameter space where supercritical propagation occurs, we describe how the expected cascade size is calculated and we describe a framework for simulating cascades using the MTBP.

\section{Complex contagion adoption dynamics \label{sec:cc_model}}

The spread of online behaviour has been shown to follow complex contagion dynamics \cite{Centola2010}. In diffusion processes on networks we say that a node has been previously exposed if one of its neighbours was active at some previous time. Here, we propose a model of complex contagion --- which extends the ICM \cite{kempe2003} --- where each additional exposure increases the probability of adoption; this is in contrast to threshold models \cite{OSullivan2015, watts2002, Unicomb2018ThresholdNetworks, monsted2017} where adoption occurs after a fraction of a node's neighbours greater than some threshold have adopted the content. De Kerchove et al.~\cite{dekerchove2009} proposed a model where adoption after one contact occurs with probability $p_{1}$ and after further contacts occurs with increased probability $p_{2}$. They assume that after the second exposure, further exposures do not increase the adoption probability. Our model builds on this to have an increase in the adoption probability after each new contact, as evidenced in Centola's work \cite{Centola2010}, for example. Here, we assume that each node is in one of three states; active, inactive or removed. At the beginning of the contagion all nodes are inactive --- they have not yet adopted --- except for the nodes selected to seed the contagion, these nodes are active in generation 0. The number of exposures that a node has received is tracked using the parameter $k$, in generation 0, $k=0$ for all nodes --- no node has received an exposure to the contagion. The parameter $\alpha,~\alpha\in[0,1]$, is used to control the strength of social reinforcement, i.e., the extent to which further exposures increase adoption probabilities. Each node that is active in generation $g$ exposes each of its neighbours, thus increasing $k$ by 1 for each neighbour, each inactive neighbour then becomes active in generation $g+1$ with its own probability $p_{k}$, otherwise they stay inactive. We define $p_{k}$ to be the probability that a node adopts given that $k$ of its neighbours have already adopted. Active nodes in generation $g$ become removed in generation $g+1$ and stay removed for the rest of the contagion. The probability that a single node will adopt after $k$ exposures, $p_{k}$, is defined as
\begin{equation}
    p_{k} = 1-q_{k} \label{eq:pn1}
\end{equation}
and
\begin{equation}
    q_{k} = q_{1}(1-\alpha)^{k-1}~, \label{eq:qn1}
\end{equation}
where $q_{k}$ is the probability of not adopting the content directly after the $k^{th}$ exposure, given that they did not adopt on any previous exposure. We can vary $\alpha$ to generate different spreading dynamics. If $\alpha = 0$, simple contagion dynamics are recovered (exposures are independent of each other) \footnote{When $\alpha = 0$ the model becomes the same as the ICM \cite{kempe2003}.}, increasing $\alpha$ increases the strength of social reinforcement in the complex contagion.\par
Equations (\ref{eq:pn1}) and (\ref{eq:qn1}) define the dynamics that we build into the MTBP model which we define shortly in Sec.~\ref{sec:mtbp}. However, we can also use network-based Monte Carlo simulations to explore the behaviour of the system. These simulations are useful for validating the MTBP model for cascades. In the network-based simulations, we begin the simulation with a randomly chosen seed node from the network which is active at the starting time. At each time step, every active node exposes all of its inactive neighbours, each becoming active at the next time step with probability $p_{k}$, active nodes become removed in the next time step. The simulation finishes when there are no active nodes left in the network. The network-based simulations are very useful for comparison; however, they can be computationally intensive and depend on an underlying network which leads to finite-size effects. In Sec.~\ref{sec:mtbp_sims} we describe a method for simulating the cascades using the MTBP framework, this method is much faster than the network-based simulations and is not subject to finite-size effects. A comparison between the speed of the network- and MTBP-based simulations is given in Table \ref{tab:simtimes}. Now that we have defined our complex contagion mechanism, in the next section, we will describe the other major component of our model, the MTBP.

\begin{figure}
    \includegraphics[width = 0.44\textwidth]{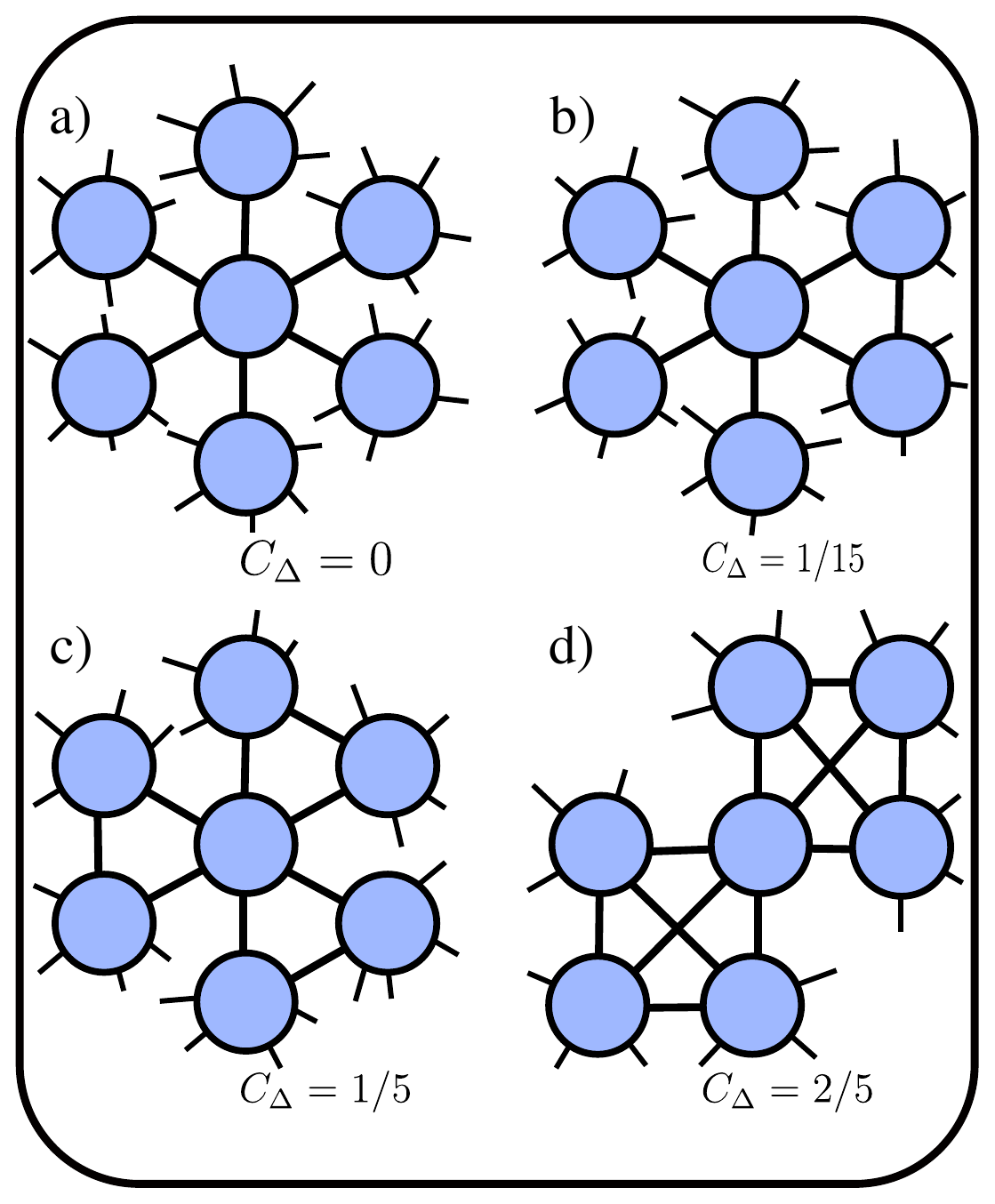}
    \caption{The local topology of each node in the networks that we study. We consider 2 classes of 6-regular networks, the first class contains networks with 3-cliques and 2-cliques only, \cite{Newman2009, Miller2009}, these are networks (a), (b) and (c) in the figure, for these networks we increase the amount of clustering by increasing the number of triangles. These are Newman-Miller networks with (a) $n_{2}=6$, $n_{3}= 0$, (b) $n_{2}=4$, $n_{3}= 1$ and (c) $n_{2}=0$, $n_{3}= 3$ as described in Sec.~\ref{sec:mtbp}. In the second class of networks which contain cliques of a specific size only (either 2, 3 or 4), these are networks (a), (c) and (d), we increase the amount of clustering by increasing the clique size. The clustering coefficient, $C_{\Delta}$, is given for each network.}
    \label{fig:ego_nets}
\end{figure}

\begin{figure*}
    \centering
    \includegraphics[width = \textwidth]{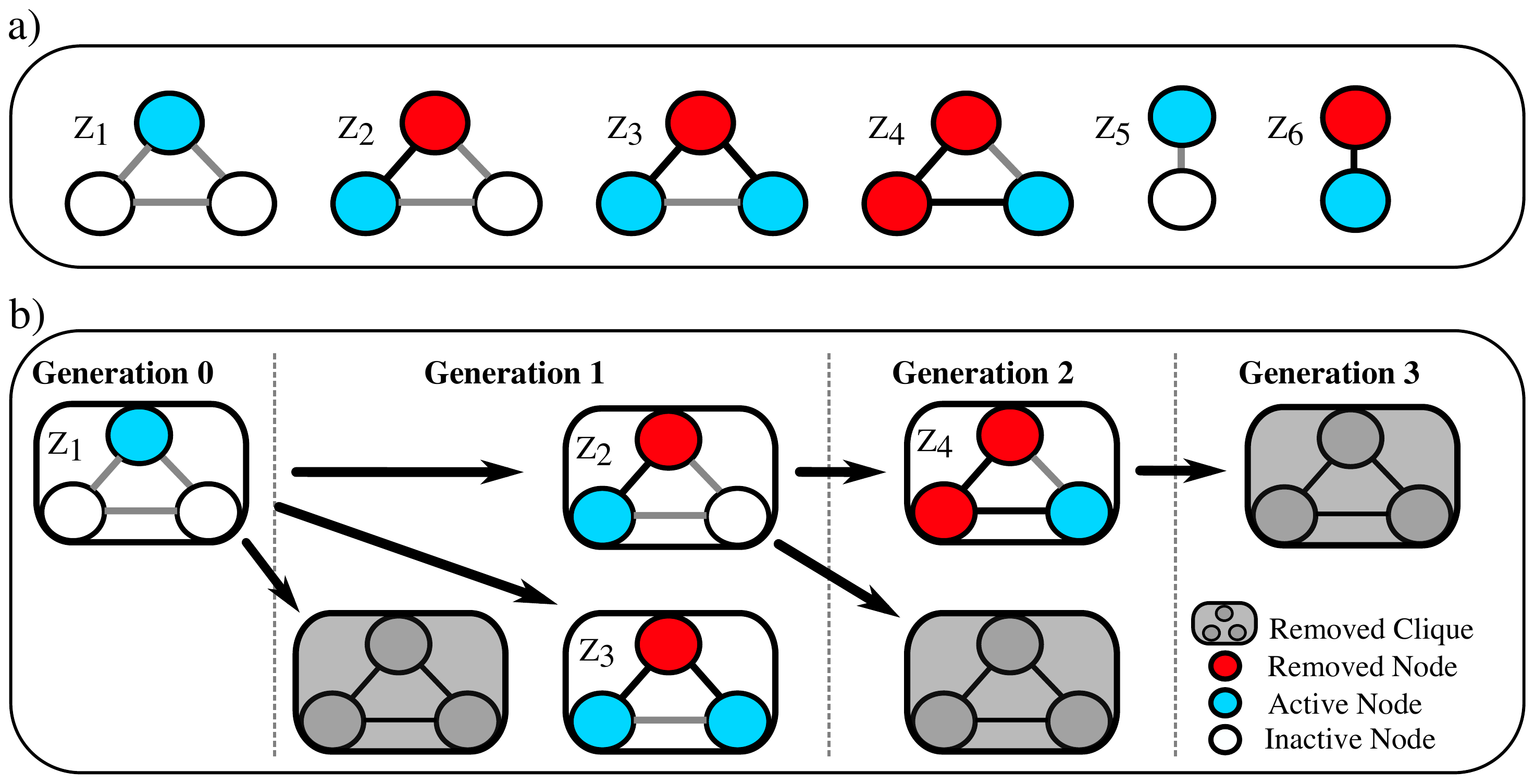}
    \caption{(a) The six motifs with at least one active node in a network composed of triangles and single edges (Newman-Miller). The blue nodes are active, the red nodes are removed and the white nodes are inactive. (b) Schematic showing the potential offspring of each motif type for a 3-clique. The motifs shaded in grey are removed as they have no active nodes. The arrows point from a motif to its potential offspring. The generation is the number of time steps since the clique first had an active node.}
    \label{fig:schematic}
\end{figure*}

\section{The Multi-type branching process model\label{sec:mtbp}}

In this section, we describe the MTBP model for a class of networks where every node in the network is part of $n_{3}$ triangles and has $n_{2}$ single edges \cite{Newman2009, Miller2009}, with $n_{3}$ and $n_{2}$ the same for each node, examples of these networks are shown in Fig.~\ref{fig:ego_nets}(a-c). We refer to networks of this class as Newman-Miller networks. Our decision to use these networks where each node has the same local structure is for ease of explanation and it is important to note that this methodology can be extended to networks where the local structure is described by clique-membership distributions. For the networks that we focus on here, different choices in seed node do not impact the dynamics. First we explain how complex contagion on this type of network can be described succinctly by a MTBP and using this description, we can readily calculate properties of the average behaviour of the cascades using a mean matrix, $\mathbf{M}$. In the second part of this section we show how we can define a mean matrix, $\textbf{M}$, for the contagion. In Appendix \ref{sec:4_cl_appendix} we describe how this methodology can be extended to 4-cliques.\par

\begin{figure}
    \centering
    \includegraphics[width = 0.45\textwidth]{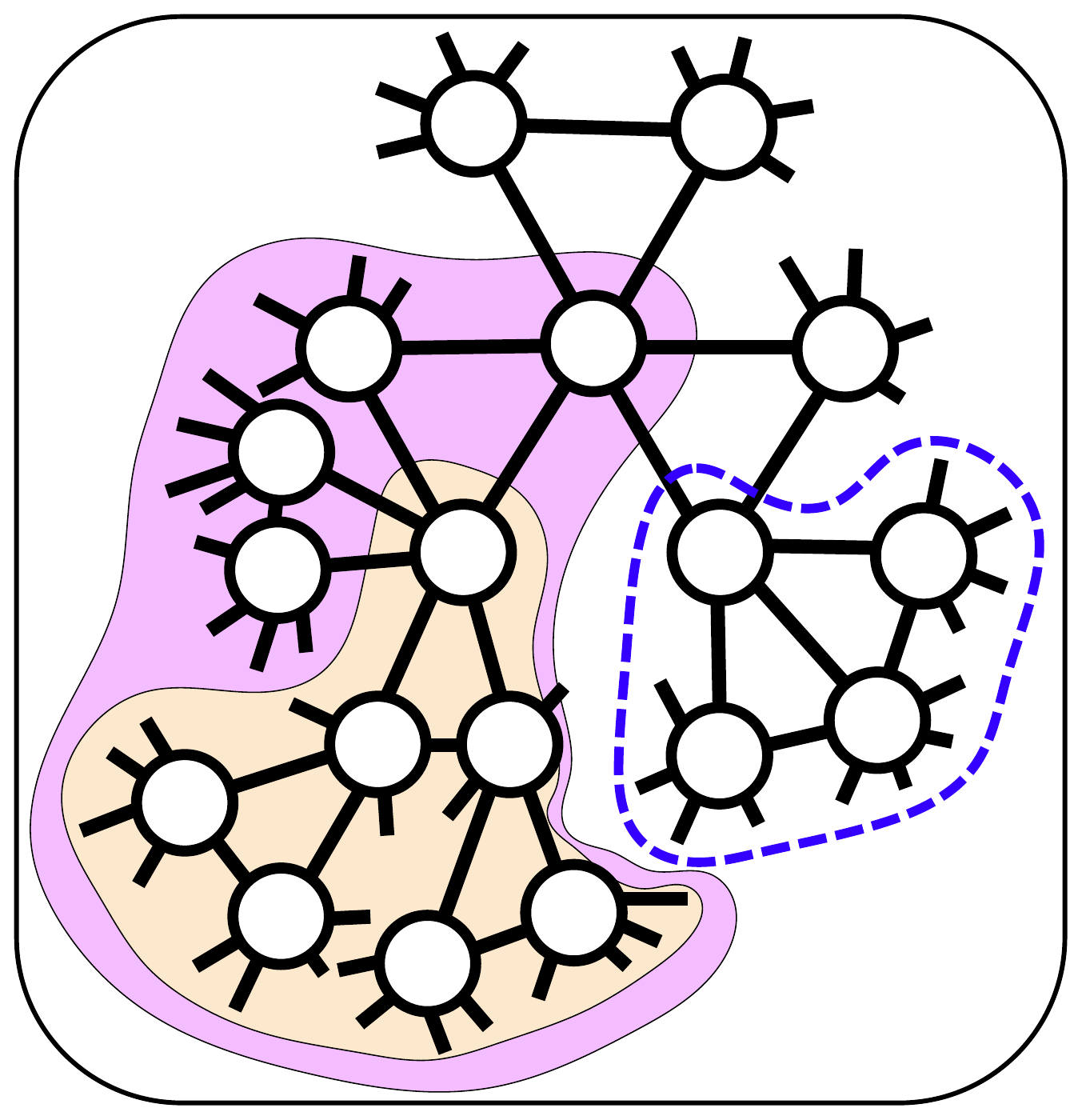}
    \caption{An illustration showing that the cliques are connected in a locally tree-like manner for the network where every node is in 3 3-cliques. If we take one branch of the network (pink) and consider a subbranch (yellow) we can see that the subbranch of the tree does not intersect with any other subbranch of the tree close to the seed node. This is what is meant by the cliques being connected in a locally tree-like manner. Inside blue-dashed region we show an example of how the finite size of a network might violate this assumption, here there are two nodes which share two 3-cliques.}
    \label{fig:tree_like}
\end{figure}

We model the spread of a complex contagion through the network using motifs. For the rest of the paper we will interchangeably refer to triangles as 3-cliques and to single links as 2-cliques. In Fig.~\ref{fig:schematic}(a) we present the six possible clique motifs of a Newman-Miller graph with $n_{3}$ 3-cliques and $n_{2}$ 2-cliques. The clique motifs can be thought of as the different adoption states that a clique can be in and we denote these states $z_{i},~i\in\{1,2,3,4\}$ for the 3-clique motifs and $i\in\{5,6\}$ for the 2-clique motifs. Each motif $z_{i}$ can produce a specific range of offspring in the next time step, each with their own probability of occurrence. In Fig.~\ref{fig:schematic}(b) the potential offspring for each type of 3-clique motif is illustrated. The potential offspring of each motif and their corresponding probabilities are shown in Table \ref{tab:offspring1}. These probabilities are derived from the complex contagion model described in Sec.~\ref{sec:cc_model}. In Table \ref{tab:offspring1} we also give the vector form of the offspring, $\vec{z}$, where the $i^{th}$ element corresponds to the number of $z_{i}$ motifs in the offspring and the binomial outcome which is the total number of new active offspring nodes from that motif. We assume that the network is so large that the probability of an active node having infected neighbours in different cliques is negligible. In these clique-based networks, by design, while each node is in $n_{3}$ 3-cliques and $n_{2}$ 2-cliques, there is no overlap between the cliques --- two cliques share at most one node. This is the same, in spirit, as the locally tree-like assumption made for mean-field approximation --- we assume that the cliques are connected in a locally tree-like manner. That is, in the construction of the network, each node is assigned to $n_{3}$ 3-cliques and $n_{2}$ 2-cliques and if the network is large enough then the probability of two nodes sharing more than one clique is low; however, for small networks this assumption may be broken and the cliques can overlap. This is illustrated in Fig.~\ref{fig:tree_like}. While the assumption that an inactive node will not have active neighbours in more than one clique is most accurate for larger networks, we show in Appendix \ref{sec:appendix_finite_size} that the MTPB methodology works quite accurately for networks with more than 100 nodes.\par
We use Table \ref{tab:offspring1} to construct the mean matrix, $\textbf{M}$, as in \cite{haccou2005}, where each entry, $m_{ij}$, $i,~j \in \{1,2,...,6\}$ of $\textbf{M}$ is the expected number of offspring motifs of type $i$ from a motif of type $j$ in the next generation. The mean matrix that we refer to here is the same as the population projection matrix (PPM) in \cite{Caswell2018MatrixModels} and is the transpose of the mean matrix in \cite{haccou2005}. There is inconsistency in the literature in the use of $\mathbf{M}$ and $\mathbf{M}^{T}$; however, as we rely closely on Caswell for our mathematical results we choose to use the version in \cite{Caswell2018MatrixModels}. The elements of the mean matrix are presented in Table \ref{tab:mij_newman}.
\begin{table*}
\begin{tabular}{|c|c|c|c|c|c|}
\hline
\textbf{Motif}           & \multicolumn{1}{c|}{\textbf{Offspring}} & \textbf{Vector}                            & \textbf{Probability} & $\mathbf{Bin(n,p)}$ & \textbf{Binomial Outcome}\\ \hline \hline
\multirow{3}{*}{$z_{1}$} & 0                                       & $(0,0,0,0,0,0)^{T}$                        & $(1-p_{1})^{2}$          & \multirow{3}{*}{$Bin(2,p_{1})$} & 0                         \\ \cline{2-4} \cline{6-6}
                         & $(n_{3}-1)z_{1}+z_{2}+n_{2}z_{5}$               & $(n_{3}-1,1,0,0,n_{2},0)^{T}$                      & $2p_{1}(1-p_{1})$    &                          &1\\ \cline{2-4} \cline{6-6}
                         & $2(n_{3}-1)z_{1+}z_{3}+2n_{2} z_{5}$            & $(2(n_{3}-1),0,1,0,2n_{2},0)^{T}$                  & $p_{1}^{2}$      &                          &2\\ \hline
\multirow{2}{*}{$z_{2}$} & 0                                       & $(0,0,0,0,0,0)^{T}$                        & $(1-p_{1})(1-\alpha)$    & \multirow{2}{*}{$Bin(1,p_{2})$} &           0              \\ \cline{2-4} \cline{6-6}
                         & $ (n_{3}-1)z_{1}+z_{4}+n_{2} z_{5}$            & $(n_{3}-1,0,0,1,n_{2},0)^{T}$                      & $\alpha +p_{1}(1-\alpha)$  &                         &1\\ \hline
\multirow{2}{*}{$z_{5}$} & 0                                       & $(0,0,0,0,0,0)^{T}$                        & $1-p_{1}$              & \multirow{2}{*}{$Bin(1,p_{1})$}  &                       0\\ \cline{2-4} \cline{6-6}
                         & $n_{3} z_{1}+(n_{2}-1)z_{5}+z_{6}$              & \multicolumn{1}{c|}{$(n_{3},0,0,0,n_{2}-1,1)^{T}$} & $p_{1}$            &                        & 1\\ \hline
\end{tabular}
\caption{The potential offspring and their corresponding vector, probability, binomial distribution and outcome for each motif in a network where each node is part of $n_{2}$ 2-cliques and $n_{3}$ 3-cliques.\label{tab:offspring1}}
\end{table*}
\begin{table}
\begin{tabular}{|c|c|}
\hline
\textbf{$(i,j)$} & \multicolumn{1}{c|}{\textbf{$m_{ij}$}} \\ \hline \hline
(1,1)            & $2(n_{3}-1)p_{1}$                      \\ \hline
(1,2)            & $(n_{3}-1)(\alpha +p_{1}(1-\alpha))$             \\ \hline
(1,5)            & $n_{3}p_{1}$                           \\ \hline
(2,1)            & $2(1-p_{1})p_{1}$                      \\ \hline
(3,1)            & $p_{1}^{2}$                        \\ \hline
(4,2)            & $\alpha +p_{1}(1-\alpha)$                  \\ \hline
(5,1)            & $2n_{2}p_{1}$                          \\ \hline
(5,2)            & $n_{2}(\alpha +p_{1}(1-\alpha))$                 \\ \hline
(5,5)            & $(n_{2}-1)p_{1}$                       \\ \hline
(6,5)            & $p_{1}$                              \\ \hline
\end{tabular}
\caption{Entries of the mean matrix $\textbf{M}$ for a network with $n_{3}$ triangles and $n_{2}$ single edges, $n_{2},n_{3}>0$, $m_{ij} = 0$ for all other $(i,j)$ combinations.\label{tab:mij_newman}}
\end{table}
In Sec. \ref{sec:results} we show how $\textbf{M}$ can be used to calculate the expected cascade size and to find the parameter regions of $(p_{1},\alpha)$ where there is supercritical diffusion.
\section{Results\label{sec:results}}
In this section we will use the multi-type branching process to generate interesting results from the complex contagion dynamics, namely, we look at the cascade condition (criticality of the cascades) and the expected cascade size. For both the cascade condition and the expected cascade size, we compare the results for 6-regular Newman-Miller networks where each node is part of $n_{3}$ triangles and $n_{2}$ single edges first. We can vary $n_{3}$ and $n_{2}$ using the rule $2n_{3}+n_{2}=6$, thus increasing $n_{3}$ results in higher levels of clustering --- by introducing more triangles --- without introducing higher order cliques. We use networks with the same local degree because it allows us to increase the amount of clustering in the network without changing other network features, thus allowing us to isolate the effect of clustering on our complex contagion spreading process. We will then compare the results for 6-regular networks with each network composed of one type of clique, in particular, we compare 6-regular networks with 2, 3 and 4-cliques. For this second class of networks, when we include larger cliques we are also increasing the clustering. We are interested in how the resulting dynamics of increasing the clustering in this way compares to the first method of including a larger proportion of triangles in the network. It is worth noting that 6-regular networks with 2-cliques only and 3-cliques only are also in the Newman-Miller class of networks where $n_{2} = 6$, $n_{3}= 0$ and $n_{2} = 0$, $n_{3}= 3$ respectively.\par
In the second part of this section we will describe a simulation technique which makes use of the MTBP methodology. These simulations are faster than usual network-based simulations and do not depend on an underlying network so do not suffer from finite-size effects, in Table \ref{tab:simtimes} we compare the simulation times for the MTBP- and network-based simulations.
\subsection{Results from the mean matrix \textbf{M} \label{sec:mean_mat}}
While doing Monte Carlo cascade simulations on networks with high levels of clustering is not any more challenging than on unclustered networks, obtaining analytic results from clustered networks cannot be achieved by many of the common methods for modelling cascade dynamics on networks which are locally tree-like. The mean matrix, $\textbf{M}$, allows us an alternative method of finding analytic results which works for clustered network structures.
\subsubsection{Cascade condition \label{sec:cascadecondition}}
\begin{figure}[ht!]
    \centering
    \begin{subfigure}{0.45\textwidth}
    \centering
\includegraphics[width = \textwidth]{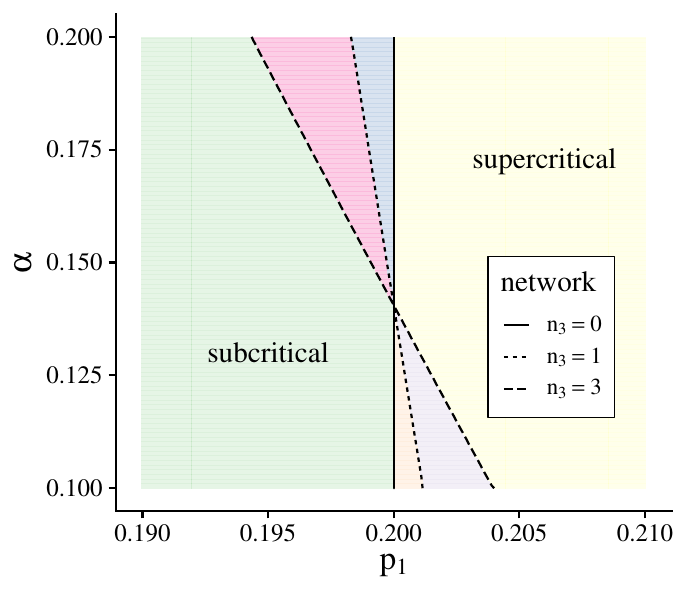}
    \caption{}
    \end{subfigure}
    \begin{subfigure}{0.45\textwidth}
    \centering
    \includegraphics[width = \textwidth]{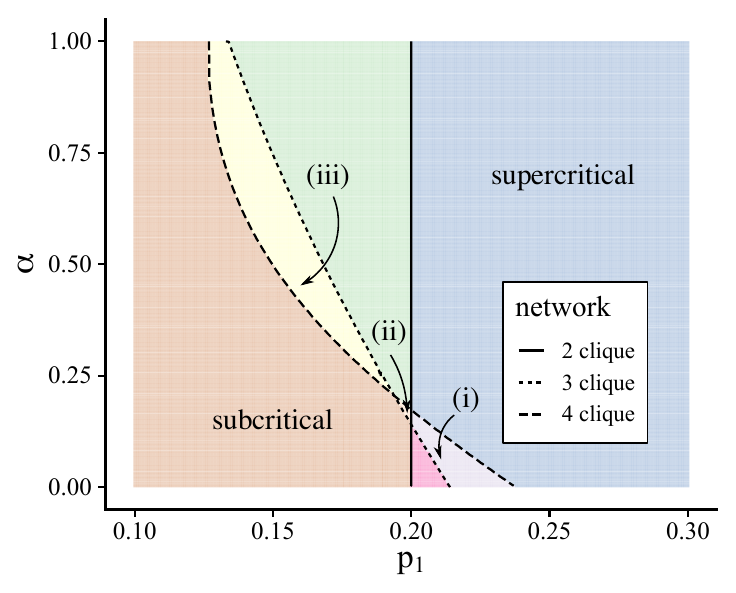}
    \caption{}
    \end{subfigure}
    \caption{Super- and subcritical parameter regions for (a) 3 different 6-regular networks where each node is part of $n_{3}$ 3-cliques and $(6 - 2n_{3})$ 2-cliques (b) 3 different 6-regular networks where each node is either part of 2 4-cliques, 3 3-cliques or 6 2-cliques. The lines represent the critical parameter boundary for each network, parameters to the left of the line produce subcritical cascades and parameters to the right of the boundary produce supercritical cascades on that network. In (b) three specific regions are labelled (i), (ii) and (iii), these are the regions where only the 2-clique, 3-clique and 4-clique networks respectively produce supercritical cascades. \label{fig:parameter_sweep}}
\end{figure}

The cascade condition determines whether an infinitesimally small seed fraction of active nodes will generate cascades of non-finite mean size, which is a non-vanishing fraction of the network as the network size tends to the large network limit. We say that the cascades are supercritical if this condition is met, otherwise, the cascades are subcritical. This is an important result because it tells us whether, on average, a social contagion is likely to die out after a few generations (subcritical) or spread to a significant portion of the network (supercritical). We use the leading eigenvalue, $\lambda_{max}$, of $\textbf{M}$ to determine the condition of the cascades. While $\textbf{M}$ does not meet the irreducibility assumption of the Perron-Frobenious theorem, in Sec.~\ref{sec:proof} we explain why the cascade condition is still described by the size of the largest eigenvalue, $\lambda_{max}$. If $\lambda_{max}>1$ the diffusion is supercritical, if $\lambda_{max}<1$ the diffusion is subcritical and if $\lambda_{max} = 1$ the diffusion is at criticality. In Fig.~\ref{fig:parameter_sweep} the $(p_{1},\alpha)$ critical boundaries are represented by lines, these transitions occur when $\lambda_{max} = 1$. For a given network, the diffusion is subcritical to the left of the line and supercritical on the right. A closed-form expression for the critical boundary for the 3-clique network is given in Appendix \ref{sec:closed_form}. The regions are obtained by calculating $\lambda_{max}$ for each parameter pair. Fig.~\ref{fig:parameter_sweep} displays the critical regions for (a) networks of $n_{3}$ 3-cliques and $n_{2}$ 2-cliques (b) networks with cliques of a single size only.\par
From these figures we see that for low levels of social reinforcement ($\alpha$ small), the critical transition occurs at larger $p_{1}$ for more clustered networks. However, once $\alpha$ is sufficiently large, the critical transition occurs at smaller $p_{1}$ on the more clustered networks. In particular, if we focus on Fig.~\ref{fig:parameter_sweep}(b), the 6-regular networks with one clique size only, we can look at specific regions of the parameter space and learn about the behaviour on each network. It is interesting here that for each of the three networks, there exists a region where there is supercritical diffusion on that network but neither of the other two. For the region labelled (i), where $\alpha$ is small and $p_{1}$ just above 0.2, there is supercritical diffusion for the 2-clique network only (clustering coefficient $C_{\Delta}=0$), for the region labelled (iii) where $\alpha$ is large and $p_{1}$ small, there is supercritical diffusion for the 4-clique network only ($C_{\Delta}=0.4$). An interesting finding is that for the region labelled (ii), where $\alpha$ and $p_{1}$ are moderate ($p_{1}$ just below 0.2) there is a region where there is supercritical diffusion on the 3-clique network only ($C_{\Delta}=0.2$), this effect is not seen in Fig.~3(a) for the networks composed of 2-cliques and 3-cliques which suggests to us that the clique size is driving this difference in the dynamics. These results are consistent with \cite{shirley2005} who show that under simple contagion dynamics $(\alpha = 0)$ clustering inhibits the diffusion on a network; however, we also see here that once $\alpha$ is large enough the opposite occurs and more clustered networks allow for supercritical diffusion at lower $p_{1}$ values. In Appendix \ref{sec:num_expt} we show that this theory agrees well with network-based simulation results.
\subsubsection{Proof that $\lambda_{max}$ determines the criticality of the cascades \label{sec:proof}}
In Sec. \ref{sec:cascadecondition}, we described how we can determine the criticality of the cascades using $\lambda_{max}$, the largest eigenvalue of the mean matrix, $\mathbf{M}$, here, we prove this result more rigorously. We begin by explaining why it works for the network where each node is part of 3 3-cliques and then discuss why the result holds in general.\par
The mean matrix, $\mathbf{M}$, for the network where each node is in 3 3-cliques has the form,
\begin{equation}
    \textbf{M} = \begin{bmatrix}
    m_{11}& m_{12}& ~0~ &~0~\\
    m_{21}&0&~0~&~0~\\
    m_{31}&0&~0~&~0~\\
    0&m_{42}&~0~&~0~
    \end{bmatrix},
\end{equation}
where $m_{ij}$ is the expected number of type-$i$ offspring from a type-$j$ motif. Note that here we do not include rows and columns that correspond to type $z_{5}$ and $z_{6}$ motifs which are presented in Table \ref{tab:mij_newman} because these motifs do not occur in networks with 3-cliques only. The number of active motifs of each type at generation $g$ are given by the vector $\vec{z}_{g}$ and the expected size of $\vec{z}_{g}$, $\langle\vec{z}_{g}\rangle$, is given by the relationship
\begin{equation}
\langle\vec{z}_{g}\rangle = \mathbf{M}\vec{z}_{g-1}.
\end{equation}
Since the matrix $\mathbf{M}$ is reducible, i.e., its associated digraph is not strongly connected, we cannot directly apply the result of the Perron-Frobenius theorem. However, $\mathbf{M}$ is in normal form, i.e., has irreducible matrices along its diagonal, and we can use the matrix in normal form to learn about the asymptotic behaviour of cascade size \cite{Caswell2018MatrixModels}. We label the diagonal blocks, $\mathbf{B}_{1},~\mathbf{B}_{2}$ and $\mathbf{B}_{3}$, in the matrix as follows;
\begin{equation}
    \textbf{M} =\left[\begin{array}{cc|c|c}
    m_{11}& m_{12}& ~0~ &~0~\\
    m_{21}&0&~0~&~0~\\ \hline
    m_{31}&0&~0~&~0~\\ \hline
    0&m_{42}&~0~&~0~
    \end{array} \right]  = \left[\begin{array}{c|c|c}
    \mathbf{B}_{1} & \begin{array}{c}
        0  \\
       0 
    \end{array} &\begin{array}{c}
        0  \\
       0 
    \end{array}\\ \hline
    \begin{array}{cc}
    m_{31}&0
    \end{array}
    &\mathbf{B}_{2}&0\\ \hline
    \begin{array}{cc}
    0&m_{42}
    \end{array}&0&\mathbf{B}_{3}
    \end{array}\right] .
    \label{eq:3cl_mean_mat}
\end{equation}
Let $S_{i}$ denote the set of motif types represented in the submatrix $\mathbf{B}_{i}$, for this particular network $S_{1} = \{z_{1}, z_{2}\}$, $S_{2} = \{z_{3}\}$ and $S_{3} = \{z_{4}\}$. The motifs in $S_{1}$ contribute to each other and may contribute to the motifs in $S_{2}$ and $S_{3}$ but do not receive contributions from $S_{2}$ and $S_{3}$. The motifs in $S_{2}$ and $S_{3}$ do not contribute to $S_{1}$. In general, if a motif is in $S_{i}$, then it can produce offspring of all types in $S_{i}$ either directly or indirectly since $\mathbf{B}_{i}$ is irreducible; however motif types in $S_{i}$ cannot directly or indirectly produce offspring in $S_{j},~j\in\{1,...,j-1\}$ since all elements above the block matrices on the diagonal are zero --- the transition rates must be zero. The elements below $B_{i}$ are not necessarily zero; therefore, it may be possible for motifs in $S_{i}$ to produce motifs in $S_{j}$, $j\in\{i+1,...,N\}$, where $N$ is the number of motif types. In the example here, the motifs in $S_{1}$ contribute to both $S_{2}$ and $S_{3}$ since $m_{31}\neq 0$ and $m_{42}\neq 0$. Because $\textbf{M}$ is reducible, the long-term dynamics depend on the initial conditions \cite{Caswell2018MatrixModels}. The states $S_{i}$ generate a set of nested invariant subspaces,
\begin{equation}
    R_{1} = S_{1} + S_{2} + S_{3},\label{eq:subspace1}
\end{equation}
\begin{equation}
    R_{2} = S_{2} + S_{3}
\end{equation}
and
\begin{equation}
    R_{3} = S_{3},\label{eq:subspace3}
\end{equation}
where $R_{3}\subset R_{2} \subset R_{1}$. A subspace $R$ is said to be invariant if $\mathbf{M}\vec{z} \in R_{i}~\forall~ \vec{z} \in R_{i}$. In other words, a trajectory that starts in $R_{i}$ but not in any of the larger subspaces will never leave $R_{i}$. In this model, we assume that the network is sparsely seeded at generation 0, this means that when the process is initiated we assume that no clique has more than one active node. Under this assumption, the trajectories can only start with motifs with one active node and all other nodes inactive. For this network the only motif of interest is $z_{1}$, this is in $S_{1}$ and thus $R_{1}$. If the trajectory begins with motifs in $S_{1}$ then it should grow at a rate given by the dominant eigenvalue of $\mathbf{M},~\lambda_{max}$ \cite{Caswell2018MatrixModels}, because a trajectory beginning in $S_{1}$ can have offspring motifs represented by all three of the irreducible matrices on the diagonal, and the set of eigenvalues of the full matrix is the same as the set of eigenvalues for all of the submatrices on the diagonal. For this specific $\mathbf{M}$, trajectories which start with motifs in $R_{2}$ do not propagate beyond generation 0, this is because the corresponding motifs do not have any inactive nodes to be activated. The eigenvalues of their corresponding blocks $\mathbf{B}_{2}$ and $\mathbf{B}_{3}$ are zero.\par
In general, for these networks we can rewrite the mean matrix in normal form by permuting the rows and columns. For example, the mean matrix for the network with $n_{2}$ 2-cliques and $n_{3}$ 3-cliques is given by 
\begin{equation} 
    \mathbf{M} =
    \begin{blockarray}{c@{\hspace{10pt}}cccccc}
    &1&2&3&4&5&6\\
    \begin{block}{c[cccccc]}
         1~&~m_{11} & m_{12}& ~0~ &~0~ &~m_{15} & ~0~   \\
         2~&~m_{21}& 0 & ~0~ &~0~ &~0 &~0~  \\
         3~&~m_{31}& 0 & ~0~ &~0~ &~0 &~0~  \\
         4~&~0 & m_{42} & ~0~ & ~0~ &~0 &~0~  \\
         5~&~m_{51} & m_{52} & ~0~ & ~0~ &~m_{55} & ~0~  \\
         6~&~0 & 0 &~0~ &~0~ & ~m_{65}&~0~\\
         \end{block}
    \end{blockarray}
\end{equation}
and by row and column permutations we get the matrix in normal form
\begin{equation}
    \mathbf{M} =  \begin{blockarray}{c@{\hspace{10pt}}cccccc}
    &1&2&5&3&4&6\\
    \begin{block}{c[ccc|c|c|c]}
         1~&~m_{11} & m_{12} &m_{15} &~0~ &~0~& ~0~   \\
         2~&~m _{21}& 0 & 0 &~0~ &~0~ &~0~  \\
         5~&~m_{51} & m_{52}  &m_{55}& ~0~ & ~0~ & ~0~  \\ \cmidrule{2-7}
          3~&~m_{31}& 0 & 0 &~0~ &~0~ &~0~  \\ \cmidrule{2-7}
         4~&~0 & m_{42} & 0 & ~0~ &~0~ &~0~  \\ \cmidrule{2-7}
         6~&~0 & 0  & m_{65}&~0~ &~0~&~0~\\
         \end{block}
    \end{blockarray}~.
\end{equation}
For the networks discussed in this paper, the first block matrix on the diagonal, $\mathbf{B}_{1}$, is always non-zero with all of the other matrices on the diagonal, $\mathbf{B}_{2},...,\mathbf{B}_{N}$, being equal to zero. For the examples used in this paper, each row under $B_{1}$ will have at least one non-zero entry, this is because the entries below $\mathbf{B}_{1}$ correspond to the expected number offspring of types which have no inactive nodes from types in $S_{1}$. The types in $S_{1}$ have at least one inactive node; thus, there a non-zero probability of all of those inactive nodes becoming active in the next time step. This means that motifs in $S_{1}$ can have offspring directly or indirectly in $S_{2},..,S_{N}$ but motifs in $S_{i}$ cannot have offspring in $S_{1},..., S_{i-1}$. To generalise Eqs.~(\ref{eq:subspace1}-\ref{eq:subspace3}), the states $S_{i}$ generate a set of invariant subspaces
\begin{equation}
    R_{1} = S_{1} + S_{2} + ... + S_{N},
\end{equation}
\begin{equation}
    R_{2} = S_{2} + ... + S_{N},
\end{equation}
\begin{equation*}
    ...
\end{equation*}
\begin{equation}
    R_{N}=S_{N}.
\end{equation}
Then as long as the trajectory begins in $R_{1}$ and none of the smaller subspaces, the long-term dynamics will be governed by the leading eigenvalue, $\lambda_{max}$, of $\mathbf{M}$ \cite{Caswell2018MatrixModels}. Practically speaking, the cascades are always seeded with motifs in $S_{1}$, because these are the only motifs that can potentially produce offspring. It is always $\lambda_{max}$ that determines whether the cascades are subcritical or supercritical.
\subsubsection{Expected cascade size}
Now that we can determine the criticality of the diffusion, we can specifically look at the subcritical regime and calculate the expected cascade size, $X$, using $\textbf{M}$. Let $x^{(g)}_{i}$ be the expected tree size of a MTBP with a $z_{i}$ seed motif after $g$ generations and $\vec{a}$ be the vector where the $i^{th}$ element, $a_{i}$, is the number of active nodes of an offspring motif infected from within the clique. For example when a $z_{1}$ motif (3-clique with one active node and 2 inactive nodes) has offspring of type $z_{2}$ (one active node, one removed node and one inactive node) it is the result of the active node in $z_{1}$ activating one of the inactive nodes in that clique; therefore, $a_{2}=1$; however, if a $z_{1}$ motif has offspring of type $z_{5}$ (a 2-clique with one active and one inactive node), it is the beginning of a new active clique, the infection did not originate inside the clique so $a_{5}=0$. We can write the recurrence relation for $x_{i}^{(g)}$
\begin{equation}
    x_{i}^{(g)}=a_{i}+\sum_{j}m_{ji}x_{j}^{(g-1)},
\end{equation}
where $m_{ij}$ is the expected number of offspring motifs of type $i$ from a motif of type $j$. This can be expressed in the form
\begin{equation}
    \vec{x}^{(g)}=\vec{a}+\textbf{M}^{T}\vec{x}^{(g-1)}.
    \label{eq:subtree_size}
\end{equation}
where $\vec{x}^{(g)}$ is the vector where the $i^{th}$ element is $x_{i}^{(g)}$. Eq.~(\ref{eq:subtree_size}) allows us to find the expected size of a subtree with seed motif of any type; however, we are interested in $X^{(g)}$ the full expected cascade size after $g$ generations. The seed node in our cascades is part of a definite set of seed motifs and thus we must treat it differently to offspring motifs for which we know the average but not the exact number of each type. We let $\vec{z}_{0}$ be the vector with $i^{th}$ entry, $z_{0i}$, equal to the number of motifs of type $i$ present at generation 0. The expected cascade size after $g$ generations, $X^{(g)}$, is calculated from the equation
\begin{equation}
    X^{(g)} = 1 + \sum_{i,j}z_{0j}m_{ij}x_{i}^{(g-1)}.\label{eq:Xnlong}
\end{equation}
Here all of the sub-trees at generation 1 of the cascade have seed particles which are the motifs represented by $\vec{z}_{0}$. Eq.~(\ref{eq:Xnlong}) can also be expressed as
\begin{equation}
    X^{(g)} = 1 + \vec{z}_{0}^{T}\textbf{M}^{T}\vec{x}^{~(g-1)}.
    \label{eq:tree_size}
\end{equation}
For subcritical cascades, the expected cascade size as $g\to\infty$ is finite and we can find $\vec{x}^{~(\infty)}$ by rearranging Eqs. (\ref{eq:subtree_size}) and (\ref{eq:tree_size}) to get
\begin{equation}
    \left( \textbf{I}-\textbf{M}^{T} \right)\vec{x}^{~(\infty)}=\vec{a}\label{eq:x_inf}
\end{equation}
and
\begin{equation}
    X^{(\infty)} = 1 +\vec{z}_{0}^{T}\textbf{M}^{T}\vec{x}^{~(\infty)},\label{eq:X_inf}
\end{equation}
where $\textbf{I}$ is the identity matrix with the same dimensions as $\textbf{M}$. We can solve Eq.~(\ref{eq:x_inf}) to find $\vec{x}^{~(\infty)}$ and then substitute into Eq.~(\ref{eq:X_inf}) to get $X^{(\infty)}$, these results are in very strong agreement with the mean cascade size of the simulations which we describe next, this is visually demonstrated in Fig.~\ref{fig:sim_comparison}(b).

\subsection{Simulation results \label{sec:mtbp_sims}}
Using the transition probabilities given in Tables \ref{tab:offspring1} and \ref{tab:offspring2} we can perform computational simulations of cascades. Simulating large ensembles of cascades in this way has the advantage of being much faster than network-based simulation methods, especially for large networks where the process can be prohibitively slow, for example, using a network with 10,000 nodes the network-based simulations are 4 times slower than the equivalent MTBP simulations. A comparison between the simulation times for network-based and branching-process type simulations is given in Table \ref{tab:simtimes}. The MTBP simulations produce cascades that correspond to the limit of infinite network size. In contrast to the results involving $\textbf{M}$ where we can study the average behaviour, using Monte Carlo simulations we can learn more about the range of cascades which can be produced from different parameter choices and study their distributions.\par
\begin{figure}
    \centering
    \includegraphics[width = 0.45\textwidth]{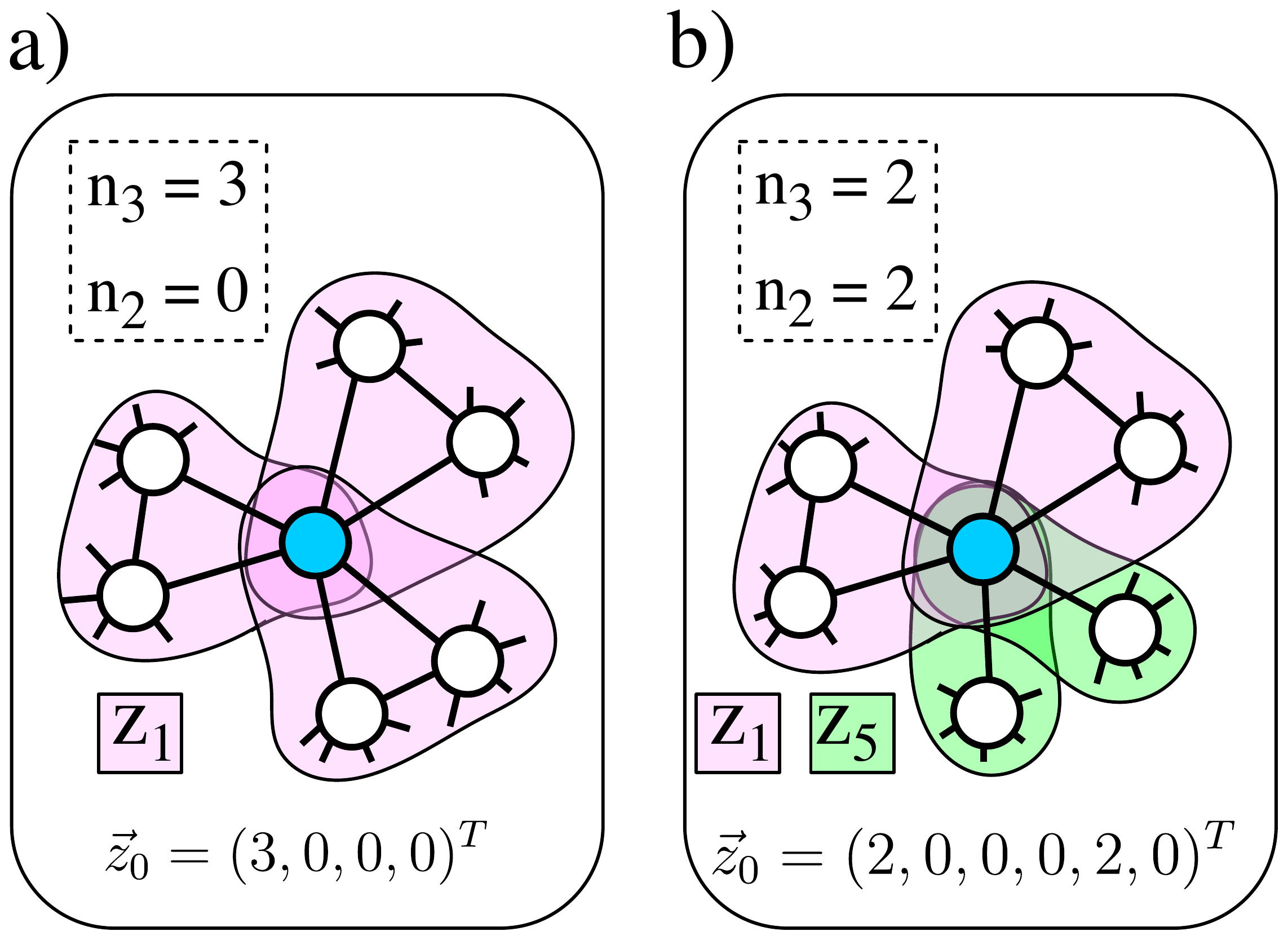}
    \caption{To construct $\vec{z}_{0}$ we examine the configuration at generation 0. We show two networks (a) the 3 3-clique network, equivalent to the Newman-Miller network with $n_{2}$ 2 cliques and $n_{3}$ 3 cliques where $n_{3} = 3,~n_{2} = 0$ and (b) the Newman-Miller network with $n_{3} = 2,~n_{2} = 2$. The initial motif vector $\vec{z}_{0}$ contains the number of motifs present of each type at generation zero, for (a) $\vec{z}_{0} = (3,0,0,0)^{T}$ and for (b) $\vec{z}_{0} = (2,0,0,0,2,0)^{T}$.}
    \label{fig:ICs}
\end{figure}

The simulation process begins with an initial distribution of motifs described by the vector $\vec{z}_{0}$ where the $i^{th}$ entry, $z_{0i}$, is the number of motifs of type $i$ at generation $g=0$. The calculation of $\vec{z}_{0}$ is illustrated in Fig.~\ref{fig:ICs}. For each further generation $g,~g>0$, each motif from $\vec{z}_{g-1}$ produces offspring from the range of possible offspring for that motif according to the binomial probabilities as described in Tables \ref{tab:offspring1} and \ref{tab:offspring2}. The simulation process finishes when $\vec{z}_{g}=\vec{0}$, i.e., when there are no new motifs produced by the previous generation.\par
\begin{figure}
\begin{subfigure}{0.45\textwidth}
    \centering
    \includegraphics[width = \textwidth]{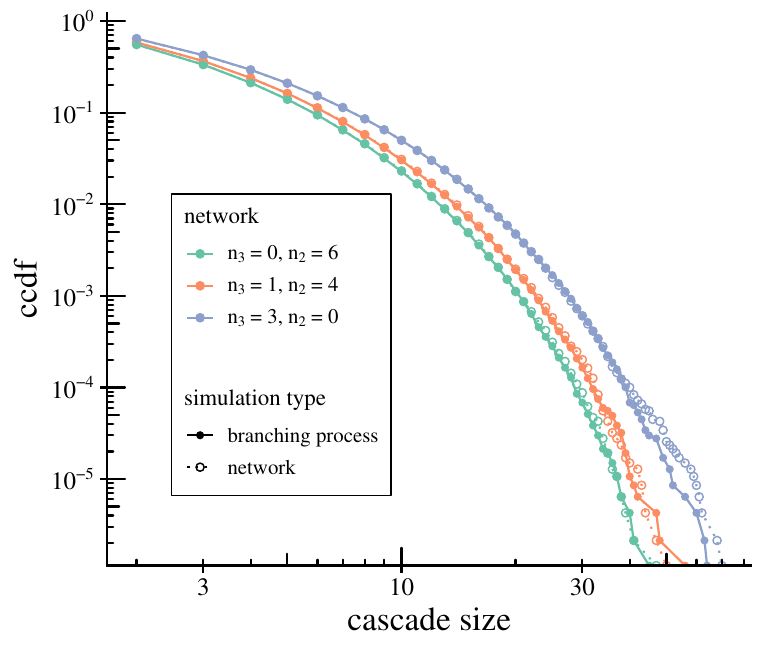}
    \caption{}
\end{subfigure}
\begin{subfigure}{0.45\textwidth}
    \centering
    \includegraphics[width = \textwidth]{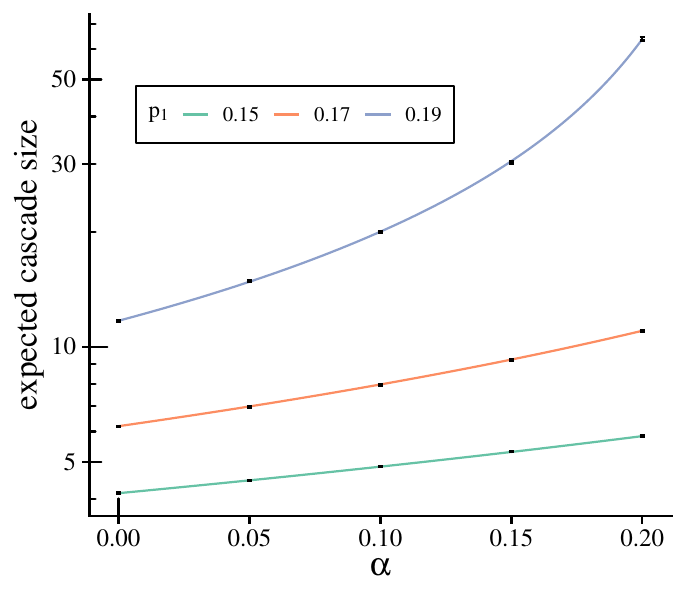}
    \caption{}
\end{subfigure}
    \caption{(a) Network-based simulations (dotted line and open circles) and MTBP simulations (solid line and points) for 6-regular networks where each node is part of $n_{3}$ triangles and $n_{2}$ single edges. The results shown are for 1 million simulations with $p_{1}=0.1$ and $\alpha = 0.2$. Networks of size 10,000 were used for the network based simulations, MTBP simulations were over 4.5 times faster for these networks. Note excellent agreement between MTBP and network-based simulations. (b) The expected cascade size for increasing $\alpha$ with $p_{1}=0.15,~0.17$ and 0.19 on networks where each node is in 3 3-cliques calculated analytically as described in Sec.~\ref{sec:mean_mat}, the error bars represent the range of values between the $2.5^{th}$ and the $97.5^{th}$ percentiles for 1 million MTBP simulations as described in Sec.~\ref{sec:mtbp_sims}. The error bars are calculated using bootstrapping, the intervals are so small that the error bars cannot be clearly seen. There is extremely strong agreement in the analytic and simulation calculations of the expected size.\label{fig:sim_comparison}}
\end{figure}

In Fig.~\ref{fig:sim_comparison}(a) we show the distribution of cascade sizes for cascades produced using this simulation scheme and for network-based simulations. We display the distributions for both simulation types on each of the networks which we discussed in Sec.~\ref{sec:mtbp}, i.e., the networks with $n_{2}$ 2-cliques and $n_{3}$ 3-cliques and single clique size networks. For the network-based simulations, we used networks with 10,000 nodes. The results from the alternate simulation methods are very similar, confirming that the behaviour in the two models is consistent. For smaller networks, we do not expect the MTBP and simulations to match up as closely. In Appendix \ref{sec:appendix_finite_size} we examine the finite size effects by simulating the cascades on smaller networks and find that while there are clear finite size effects for very small networks, the MTBP simulations compare quite closely to the network-based simulations on networks of different sizes.
\begin{table}[h]
\begin{tabular}{|c|c|c|}
\hline
Network Size & Simulation Time (s) & Ratio \\ \hline
100          & 81                  & 1.8      \\
1,000        & 91                  & 2.1      \\
10,000       & 203                 & 4.6      \\ \hline
\end{tabular}
\caption{Simulation time for 100,000 network-based simulations on networks where each node is in 3 3-cliques. Times are given for networks of size 100, 1,000 and 10,000 and compared to the time taken to do 100,000 branching-process type simulations which takes 43.9 seconds. The ratio between the network-based and branching-process simulation times is given in the table, from this we can compare the simulation times of the MTBP- and network-based methods. Simulations were conducted in R version 3.6.1 on a Quad-Core Intel Core i5 processor.\label{tab:simtimes}}
\end{table}
\section{Conclusions}
In simple branching processes, it is not possible to incorporate complex contagion dynamics because social reinforcement coming from nodes other than the parent node cannot be easily accounted for. Therefore, to model complex contagion on clustered networks, we have to go beyond simple branching processes and use Multi-Type Branching Processes (MTBPs) to model the contagion. There are three main contributions from this paper; a simple extension of the Independent Cascade Model (ICM) to a complex contagion for the spread of information, the introduction of the  mean matrix $\textbf{M}$ which we use to study the average behaviour of the system and a computationally efficient method for simulating cascades using the MTBP methodology.\par
In the adoption function introduced in Eqs.~(\ref{eq:pn1}) and (\ref{eq:qn1}) the more previous exposures that an inactive node has, the higher the probability that they will become activated by an active node. The increase in activation probability decreases as the number of exposures gets large, this type of behaviour was empirically observed by Centola \cite{Centola2010}. This model is different to the threshold models which are typically used to model complex contagion and captures empirically observed behaviour, which is not captured by threshold models \cite{Centola2010}. Its dependence on two parameters enables it to be reduced to the ICM in one natural limit, while yielding non-trivial structures in the two-dimensional parameter space.\par
The main focus of this paper was the use of MTBPs to model diffusion on clustered networks. MTBP methods have been used to model populations in ecology since the 1940s \cite{Caswell2018MatrixModels} and we adapted these established methods to model cascades on networks with different amounts of clustering and clique sizes. We took care to use networks with homogenous local structure so as to isolate the impact of varying the amount of clustering and the clique size, these methods can be extended to networks with inhomogenous local structure. While in ecology the offspring particles are usually new organisms being born, in the extension to cascades of information on social networks, the offspring particles are clique motifs with varying numbers of active, inactive and removed nodes. The MTBP formulation of the dynamics allowed us to define the mean matrix, $\mathbf{M}$, which describes the average behaviour of the system. Using $\textbf{M}$, a range of interesting analytical results can be derived, here we showed how to find the parameter regions where the cascades are subcritical (and supercritical) and how to calculate the expected cascade size. The range of possible analytical results is by no means restricted to what we have shown here, for more examples of what results are possible using MTBPs see Caswell \cite{Caswell2018MatrixModels} and Haccou et al.~\cite{haccou2005}.\par
The MTBP framework also lends itself to a method for simulating cascades. While from $\textbf{M}$ we obtain results about the average behaviour of the cascades, using the simulation strategy we can learn about the range of behaviour of the cascades. These simulations are much faster than network-based methods and assume an infinitely large network with homogenous local topology. Thus, they are not affected by finite size effects. We compared the multi-type branching process simulations to network-based simulations on a large network (10,000 nodes) in Fig.~\ref{fig:sim_comparison} and observed very strong agreement in the results.\par
The theory described here for modelling complex contagion on clustered networks has potential to be extended to a much wider range of real-world networks. It is possible to extend the methods described here to networks with more general degree distributions and clique-membership distributions. A possible direction for future work is to develop a method for approximating real-network distributions using clique distributions, this would enable the method to be extended to real-world networks. The analysis here was restricted to spreading processes where active nodes become immediately and certainly removed in the next time step, a future direction for this research would be to look at asynchronous spreading processes in a MTBP framework. Real world spreading processes are also dependent on external factors, for example the limited attention that a user has available to distribute amongst their friends \cite{lerman2016information} and spontaneous adoption due to influence outside of the network being studied \cite{ruan2015}, these are other possible ways in which the model could be extended. We anticipate that the framework described here will assist in the understanding of the dynamics of complex contagion under different levels of reinforcement and adoption probabilities and on networks with different topologies.
\begin{acknowledgements}
This work was supported by Science Foundation
Ireland [grant numbers 18/CRT/6049 (L.K.), 16/IA/4470 (J.G.), 16/RC/3918 (J.G.), 12/RC/2289 P2 (J.G.)]  with co-funding from the European Regional Development Fund.
\end{acknowledgements}
\appendix
\section{Mean matrix for 4-clique network\label{sec:4_cl_appendix}}
\begin{figure}
    \centering
    \includegraphics[width = 0.48\textwidth]{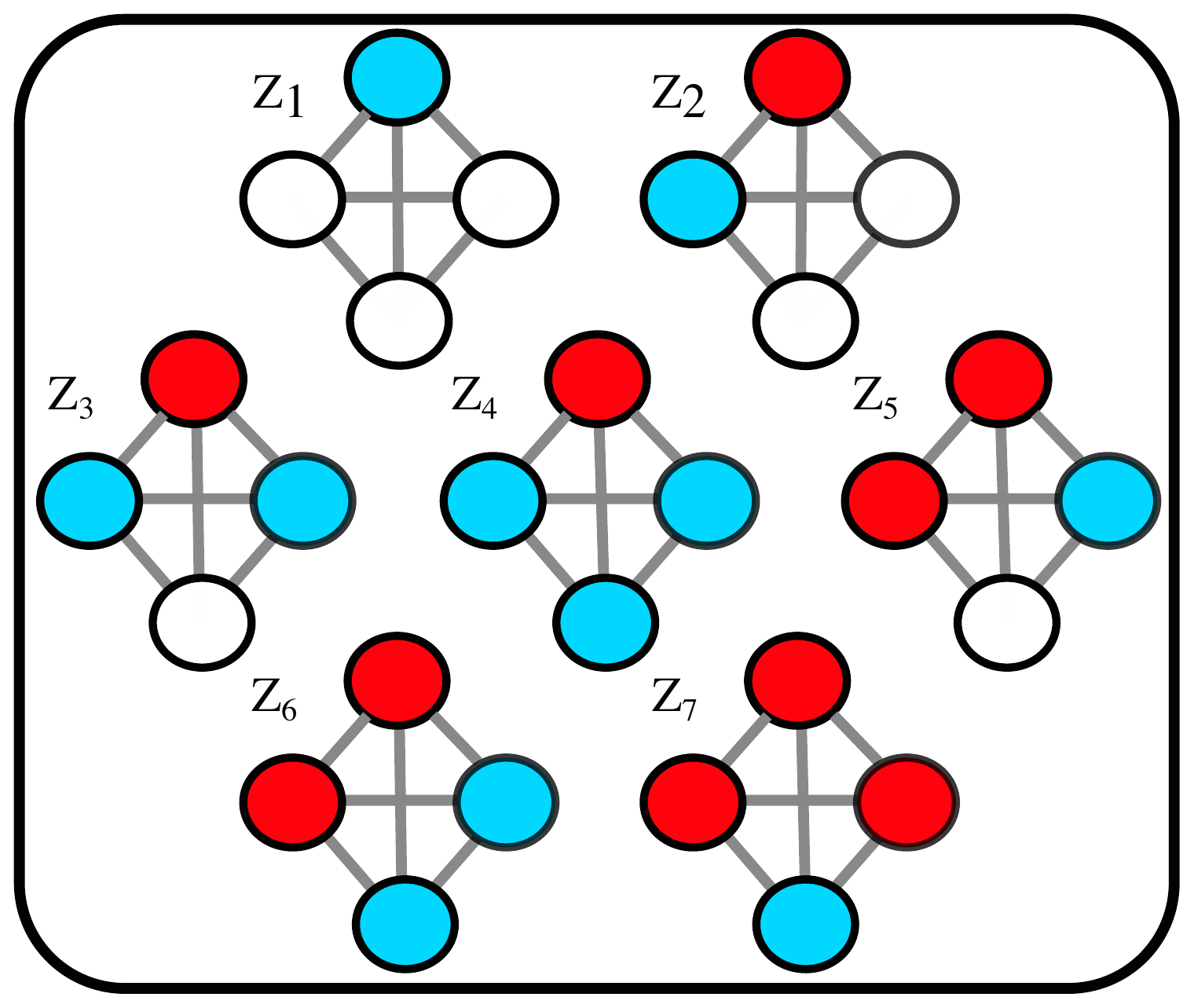}
    \caption{The seven possible motifs of a 4-clique. The blue nodes are active, the red nodes are removed and the white nodes are susceptible.}
    \label{fig:4cl_motifs}
\end{figure}

In Sec. \ref{sec:mtbp} we showed how to construct $\textbf{M}$ for a network composed of 3-cliques and 2-cliques, here we demonstrate how higher-order cliques can be included in the model. We focus on the example of a network where every node is part of $n_{4}$ 4-cliques, shown in Fig.~\ref{fig:ego_nets}(d). For the example described in Sec. \ref{sec:mtbp} each motif had at most 1 active node which could induce other nodes in the clique to become active in the next time step; however, in the case of 4-cliques, two nodes can be simultaneously active in a motif with at least one inactive node, as shown in Fig.~\ref{fig:4cl_motifs}, this requires us to be more careful with the offspring probability calculations. To explain this in more detail, assume that we have one active node, then each inactive node becomes active in the next time step with probability $p_{k}$, where $k$ is the total exposures that a node has received prior to becoming active. However, if there are two active nodes, then the first active node will expose each of its inactive neighbours and simultaneously attempt to activate each with probability $p_{k}$, if the activation is unsuccessful, the second active node then exposes the inactive node again and activates it with probability $p_{k+1}$. If an inactive node has had $k-1$ exposures before the current time step then the probability of it being activated if there are exactly two active neighbours is
\begin{equation}
    1-q_{k}q_{k+1}~,
\end{equation}
where $q_{k}=1-p_{k}$ as in Eq.~(\ref{eq:pn1}). Using this, we construct Table \ref{tab:offspring2} which shows the offspring of each motif with the corresponding probabilities, offspring vectors and binomial outcomes similarly to Sec. \ref{sec:mtbp} and use these probabilities to calculate the elements, $m_{ij}$, of $\textbf{M}$ for the 4-clique network which are shown in Table \ref{tab:mij_clique}.
\begin{table*}
\begin{tabular}{|c|c|c|c|c|}
\hline
\textbf{Motif}           & \multicolumn{1}{c|}{\textbf{Offspring}} & \textbf{Vector}                            & \textbf{Probability} & \textbf{Binomial Outcome} \\ \hline \hline
\multirow{4}{*}{$z_{1}$} & 0                                       & $(0,0,0,0,0,0,0)^{T}$                        & $(1-p_{1})^{3}$          & 0                         \\ \cline{2-5} 
                         & $(n_{4}-1)z_{1}+z_{2}$               & $(n_{4}-1,1,0,0,0,0,0)^{T}$                      & $3(1-p_{1})^{2}p_{1}$    & 1                         \\ \cline{2-5} 
                         & $2(n_{4}-1)z_{1}+z_{3}$            & $(2(n_{4}-1),0,1,0,0,0,0)^{T}$                  & $3(1-p_{1})p_{1}^{2}$      & 2 
                         \\ \cline{2-5}
                         & $3(n_{4}-1)z_{1}+z_{4}$            & $(3(n_{4}-1),0,0,1,0,0,0)^{T}$                  & $p_{1}^{3}$      & 3                         \\ \hline
\multirow{3}{*}{$z_{2}$} & 0                                       & $(0,0,0,0,0,0,0)^{T}$                        & $(1-p_{1})^{2}(1-\alpha)^{2}$    & 0                         \\ \cline{2-5} 
                         & $(n_{4}-1)z_{1}+z_{5}$            & $(n_{4}-1,0,0,0,1,0,0)^{T}$                      & $2 (1-p_{1})(1-\alpha)(\alpha+p_{1}(1-\alpha))$  & 1 
                         \\ \cline{2-5}
                         & $2(n_{4}-1)z_{1}+z_{6}$                                       & $(2(n_{4}-1),0,0,0,0,0,1,0)^{T}$                        & $(\alpha+p_{1}(1-\alpha))^{2}$    & 2                         \\ \hline
\multirow{2}{*}{$z_{3}$} & 0                                       & $(0,0,0,0,0,0,0)^{T}$                        & $(1-p_{1})^{2}(1-\alpha)^{3}$              & 0                         \\ \cline{2-5} 
                         & $(n_{4}-1)z_{1}+z_{7}$              & \multicolumn{1}{c|}{$(n_{4}-1,0,0,0,0,0,1)^{T}$} & $1-(1-p_{1})^{2}(1-\alpha)^{3}$            & 1                         \\ \hline
\multirow{2}{*}{$z_{5}$} & 0                                       & $(0,0,0,0,0,0,0)^{T}$                        & $(1-p_{1})(1-\alpha)^{2}$              & 0                         \\ \cline{2-5} 
                         & $(n_{4}-1)z_{1}+z_{7}$              & \multicolumn{1}{c|}{$(n_{4}-1,0,0,0,0,0,1)^{T}$} & $1-(1-p_{1})(1-\alpha)^{2}$            & 1                         \\ \hline                         
\end{tabular}
\caption{The potential offspring and their corresponding vector, probability and binomial outcome for each motif in a network where each node is part of $n_{4}$ 4-cliques.\label{tab:offspring2}}
\end{table*}
\begin{table}
\begin{tabular}{|c|c|}
\hline
\textbf{$(i,j)$} & \multicolumn{1}{c|}{\textbf{$m_{ij}$}} \\ \hline \hline
(1,1)            & $3(n_{4}-1)p_{1}$                      \\ \hline
(1,2)            & $2(n_{4}-1)(\alpha+p_{1}(1-\alpha))$            \\ \hline
(1,3)            & $(n_{4}-1)(1-(1-p_{1})^{2}(1-\alpha)^{3})$     \\ \hline
(1,5)            & $(n_{4}-1)(1-(1-p_{1})(1-\alpha)^{2})$         \\ \hline
(2,1)            & $3(1-p_{1})^{2}p_{1}$                  \\ \hline
(3,1)            & $3(1-p_{1})p_{1}^{2}$                  \\ \hline
(4,1)            & $p_{1}^{3}$                        \\ \hline
(5,2)            & $2(1-p_{1})(1-\alpha)(\alpha+p_{1}(1-\alpha))$  \\ \hline
(6,2)            & $(\alpha+p_{1}(1-\alpha))^{2}$              \\ \hline
(7,3)            & $1-(1-p_{1})^{2}(1-\alpha)^{3}$            \\ \hline
(7,5)            & $1-(1-p_{1})(1-\alpha)^{2}$                \\ \hline
\end{tabular}
\caption{Entries of the mean matrix $\textbf{M}$ for a network with $n_{4}$ 4 cliques, $n_{4}>0$, $m_{ij} = 0$ for all other $(i,j)$ combinations, where $m_{ij}$ is the expected number of type $i$ offspring from a type $j$ motif.\label{tab:mij_clique}}
\end{table}
\section{Network size limitations \label{sec:appendix_finite_size}}
\begin{figure}[h!]
    \centering
    \includegraphics[width = 0.48\textwidth]{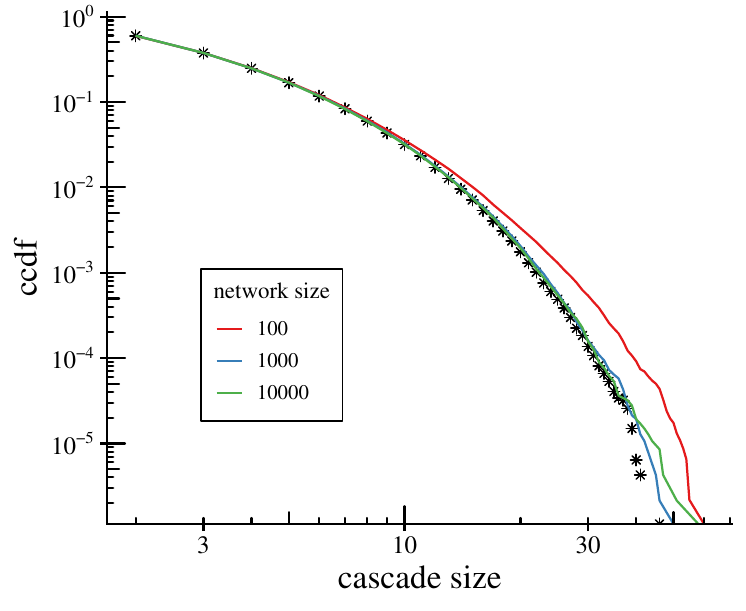}
    \caption{The ccdf for the cascade size for network-based simulations on networks of sizes 100, 1,000 and 10,000 (lines) and branching process simulations (points). Distributions shown are for 1 million simulations each with $\alpha=0.1$ and $p_{1}=0.1$ on networks where each node is part of 3 3-cliques.}
    \label{fig:finite_size}
\end{figure}
The MTBP framework assumes an infinitely large network; therefore, random exposures from other parts of the network are not considered. However, real networks are of finite size and as network size decreases, the probability that a node receives additional exposures from other parts of the network increases. In Fig.~\ref{fig:finite_size} we compare cascade size distributions for simulations on networks of sizes 100, 1,000 and 10,000 to the cascade size distribution from the MTBP simulation scheme. The MTBP type simulation distribution is very close to the results for networks of sizes 1,000 and 10,000 but for the network with 100 nodes we observe a larger proportion of bigger cascades than would be expected from the MTBP framework. This is due to nodes having a greater chance of receiving additional exposures from nodes in other parts of the network and in the context of our approach can be seen as a finite size effect.

\section{Closed-form expression for critical $\alpha$ for the 3-clique case \label{sec:closed_form}}

The cascades are said to be at criticality when the largest eigenvalue of $\mathbf{M}$, $\lambda_{max}$, is 1, as described in Sec.~\ref{sec:results}. If we are to have a closed-form expression for the relationship between the parameters $p_{1}$ and $\alpha$ at criticality, we need a closed-form expression for $\lambda_{max}$ in terms of these parameters. Finding the eigenvalues of a $n\times n$ matrix involves solving a degree-$n$ polynomial; however, polynomials in general form of degree 5 or more cannot be solved algebraically according to Abel's impossibility theorem \cite{Abel2012DemonstrationDegre}. The only network described in this paper for which we can get a closed-form solution for $\lambda_{max}$ is the network composed of 3-cliques only, as shown in Fig.~\ref{fig:ego_nets}(c) which has a $4\times 4$ mean matrix. The matrix is shown in Eq.~(\ref{eq:3cl_mean_mat}). The largest eigenvalue of this matrix is given by
\begin{equation}
    \lambda_{max} = \frac{1}{2}\left(m_{11}+\sqrt{m_{11}^{2}+4m_{12}m_{21}}\right),
\end{equation}
where $m_{11},~m_{12}$ and $m_{21}$ are defined in Table \ref{tab:mij_newman}. If we let $\lambda_{max} = 1$ then we can solve for the critical social reinforcement value, $\alpha^{*}$, in terms of $p_{1}$ and the number of 3-cliques per node, $n_{3}$. We find that 
\begin{equation}
    \alpha^{*} = \frac{2(n_{3}-1)(-p_{1}^{3}+p_{1}^{2}+p_{1})-1}{2(n_{3}-1)(-p_{1}^{3}+3p_{1}^{2}-2p_{1})}.
\end{equation}
If $\alpha>\alpha^{*}(p_{1})$ then the cascades will be supercritical. If we examine the dependence of $\alpha^{*}$ on $p_{1}$, we can see that as $p_{1}$ increases, $\alpha^{*}$ decreases up the critical value for $p_{1}$ when $\alpha = 0$, i.e., $p_{1}^{*}$ for a simple contagion. The larger $p_{1}$ is, the smaller the corresponding value for $\alpha^{*}$ will be.

\section{Agreement between theoretical threshold and simulations \label{sec:num_expt}}
\begin{figure}
    \centering
    \includegraphics[width = 0.45\textwidth]{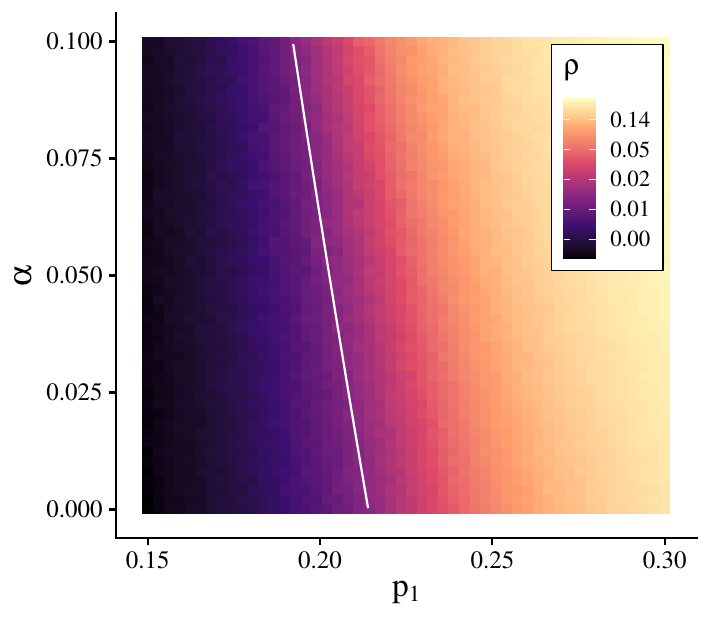}
    \caption{Mean fraction of network involved in the outbreak, $\rho$, for varying $(p_{1},\alpha)$ parameter combinations. The white line represents the theoretical threshold in the parameter space for supercritical and subcritical diffusion as described in Sec.~\ref{sec:cascadecondition} for which a closed-form expression is given in Appendix \ref{sec:closed_form}.\label{fig:num_expt}}
\end{figure}
To verify the result in Sec.~\ref{sec:cascadecondition} we ran network-based simulations on a network where each node is in 3 3-cliques and calculated the mean fraction of nodes in a cascade, $\rho$. For subcritical regions we expect $\rho$ to be very small and for supercritical regions we expect $\rho$ to be a substantial proportion of the network. In Fig.~\ref{fig:num_expt} $\rho$ is plotted against varying $p_{1}$ and $\alpha$. The critical threshold from Appendix \ref{sec:closed_form} is also shown and we can see strong agreement between the theoretical prediction and the numerical results.
\FloatBarrier
\bibliography{references}

\end{document}